%% file: main.tex
\documentclass{vldb}
\usepackage{graphicx}
\usepackage{balance}  
\usepackage{enumitem}

\usepackage{caption} 
\usepackage{balance}       
\usepackage{graphics}      
\usepackage[T1]{fontenc}   
\usepackage{txfonts}
\usepackage{mathptmx}
\usepackage[pdflang={en-US},pdftex]{hyperref}
\usepackage{color}
\usepackage{booktabs}
\usepackage{textcomp}
\usepackage[linesnumbered,ruled]{algorithm2e}
\usepackage[noend]{algpseudocode}
\usepackage{algorithmicx,algpseudocode}
\usepackage{ifthen}
\usepackage{graphicx}
\usepackage{epsfig}
\usepackage{amssymb}
\usepackage{subcaption}
\usepackage{bm}
\usepackage{url}

\usepackage{tabularx,multicol,multirow}
\usepackage[dvipsnames,table]{xcolor}

\begin{document}


\title{Annotation of Car Trajectories based on Driving Patterns}



%
%
%
%

\numberofauthors{3} 

\author{Sobhan Moosavi$^{\dag}$, Behrooz Omidvar-Tehrani$^{\dag}$, R. Bruce Craig$^{\ddag}$, Rajiv Ramnath$^{\dag}$\\
{$^{\dag}$CSE Department, Ohio State University; $^{\ddag}$Nationwide Insurance}\\
{$^{\dag}$\{moosavinejaddaryakenari.1,omidvar-tehrani.1,ramnath.6\}@osu.edu; $^{\ddag}$craigr2@nationwide.com}
\bigskip
}

\maketitle

\begin{abstract}
Nowadays, the ubiquity of various sensors enables the collection of voluminous datasets of car trajectories. Such datasets enable analysts to make sense of driving patterns and behaviors: in order to understand the {\em behavior} of drivers, one approach is to break a trajectory into its underlying {\em patterns} and then analyze that trajectory in terms of derived patterns. The process of trajectory segmentation is a function of various resources including a set of ground truth trajectories with their driving patterns. To the best of our knowledge, no such ground-truth dataset exists in the literature. In this paper, we describe a trajectory annotation framework and report our results to annotate a dataset of personal car trajectories.
Our annotation methodology consists of a crowd-sourcing task followed by a precise process of aggregation. Our annotation process consists of two granularity levels, one to specify the annotation (segment border) and the other one to describe the type of the segment (e.g. speed-up, turn, merge, etc.). 
The output of our project, Dataset of Annotated Car Trajectories (DACT), is available online at \url{https://figshare.com/articles/dact\_dataset\_of\_annotated\_car\_trajectories/5005289}.
\end{abstract}

\input{Introduction}
\input{ProbState}
\input{TrajectoryAnnotation}
\input{Dataset}
\input{Results}
\input{Output}
\input{Conclusion}
\input{Acknowledgement}

\bibliographystyle{abbrv}
\bibliography{references}



\end{document}

%% file: Introduction.tex
\section{Introduction}
\label{sec:intro}

Nowadays, the ubiquity of various sensors (accelerometer, barometer, GPS, etc.) enables the collection of voluminous datasets of car trajectories. Examples of such datasets are the New York taxi cab \cite{nycTaxi}, Porto taxi cab \cite{moreira2013predicting}, and GeoLife \cite{geolife-gps-trajectory-dataset-user-guide}. Such datasets enable analysts to make sense of driving patterns and behaviors. By exploring driver behaviors and comparing different drivers, an analyst provides solutions to increase drivers safety and decrease the risk.\footnote{\scriptsize The risk can be interpreted from an insurance perspective.}

\begin{figure}[t]
  \includegraphics[scale=0.34]{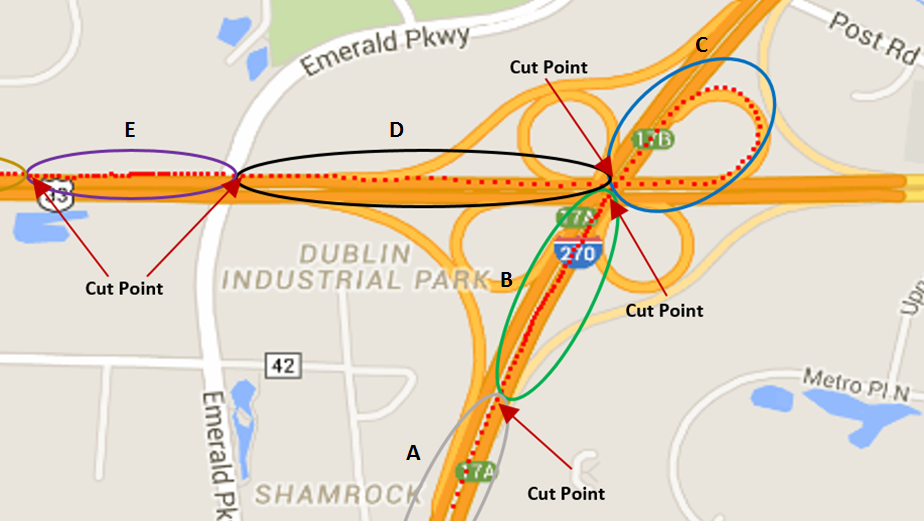}
  \caption{A sample trajectory with several underlying driving patterns specified by ovals. Also, red arrows show the points of transition between patterns, where new patterns commence.}
  \label{fig:patterns}
\end{figure}

We consider a dataset of car trajectories where each of trajectory is a time-stamped sequence of data points and each data point is described using attributes such as speed, bearing, location, etc. 
In order to understand driver behavior, one approach is to break each trajectory into its underlying patterns, i.e., segmentation, and then describe the trajectory in terms of derived patterns. One example of a trajectory and its underlying patterns is depicted in Figure \ref{fig:patterns}. 

Trajectory segmentation has been thoroughly studied in the literature. Buchin et al. propose to exploit spatiotemporal criteria (like changes in speed or direction of moving vehicle) for segmentation \cite{buchin2011segmenting}. Explorations of multiple settings for stable criteria in terms of monotonic and non-monotonic attributes is another approach which is introduced by Alewijnse et al. \cite{alewijnse2014framework}. Moreover, there are several other research tracks which aim to solve the segmentation problem by statistical modelings and other algorithmic approaches. For instance Alewijnse et al. \cite{alewijnse2014model} propose to use Brownian Bridge Model to model the movement and leverage a dynamic programming approach to discover segments.
The aforementioned approaches are mostly focused on animal movement trajectories and do not discuss human behavior.  
However, there exists few generic approaches such as the seminal work by Anagnostopoulos et al. which leverages a geometric based approach for segmenting human generated trajectories \cite{anagnostopoulos2006global}. Also in a recent study, we  proposed an approach to transform a trajectory based on a statistical modeling and then applied a dynamic programming-based segmentation approach and identified the best number of segments by using Minimum Description Length (MDL) \cite{moosavi2016discovery}. 

One of the main shortcomings of all aforementioned efforts is that the applicability and performance of the approach is often exhibited through formal statements or a few real-world use cases. This does not provide the evaluation of the underlying algorithms in practice. For such evaluation, there should be a set of labeled trajectories as ground truth, where the true segment borders are specified in trajectories by experts\footnote{\scriptsize Such task can be seen as a domain-specific task, where each domain requires its own expert for annotation.}. However, to the best of our knowledge, there is no such dataset available providing this information. 

In this paper we describe a novel {\em trajectory annotation framework} and present the methodology of our annotation process comprising two main steps: {\em Expert-Annotation} and {\em Annotation-Aggregation}. Moreover, we report the result of using the proposed framework to annotate a dataset of personal car trajectories. For annotation purpose, we ran a crowd-sourcing task followed by a precise process of aggregation. 

The rest of this paper is organized as follows: First, we formally introduce the problem annotating car trajectories in Section \ref{sec:prob}. Next, trajectory annotation framework is discussed in details in Section \ref{sec:annotation}, followed by a description of the trajectory dataset in Section \ref{sec:data}. Section \ref{sec:res} provides annotation results. Finally, we conclude in Section \ref{sec:conc}.

%% file: ProbState.tex
\section{Problem Statement}
\label{sec:prob}
We are given a trajectory dataset $T = \{T_1, T_2, \dots, T_N \}$ of size $N$, where each trajectory $T_i$, $1 \leq i \leq N$, is a sequence of data points $\langle p_{i1}, p_{i2}, \dots, p_{in} \rangle$. Also, each data point $p_{ij}$, $1 \leq j \leq n$, is a tuple of the form $(Time_j, Speed_j, Heading_j, Lat_j$ $,Lng_j)$. In this tuple, $Time_j$ is the time stamp, $Speed_j$ shows the vehicle's speed (mph), $Heading_j$ shows the change in direction of moving vehicle based on the direction in previous time stamp, and $Lat_j$ and $Lng_j$ represent the exact location of vehicle at the current time stamp. 

\newtheorem{mydef}{Definition}
\begin{mydef}
\label{def:seg}
Segmentation of a trajectory $T$ is the task of breaking $T$ to a set of non-overlapping sequences of sub-trajectories. 
\end{mydef}

In Definition \ref{def:seg}, the segmentation of a trajectory $T=\langle p_{i1}, p_{i2},$ $\dots, p_{in} \rangle$ into $m$ segments, is about finding a set of cutting indexes $\langle I_1, I_2$ $\dots, I_m \rangle$ that mark the end points of the segments in $T$. Thus, we can define a set of cutting data points for the segmented trajectory $T$ as $\langle p_{I_1}, p_{I_2} \dots, p_{I_m} \rangle$. Note that $p_{I_m}=p_n$. All data points between indexes $I_i$ and $I_{i+1}$, excluding point $\rho_{I_i}$ and including point $\rho_{I_{i+1}}$, belong to the $(i+1)^{th}$ segment.
An example of a segmented trajectory with all segments and cutting points is shown in Figure \ref{fig:patterns}. 

\begin{mydef}
Annotation of a trajectory $T$ is the task of identification of segment borders (cut points) within $T$ using human expertise. Moreover, an expert can assign one or more labels (segment types) to each segment.
\end{mydef}

A label can be a description of the pattern which occurs in a segment. For example, segment $B$ in Figure \ref{fig:patterns} can be labeled as {\em slow-down} and segment $C$ as {\em loop}. 

%% file: TrajectoryAnnotation.tex
\section{Annotation Framework}
\label{sec:annotation}

We propose a novel framework for trajectory annotation. This framework has two important components: $i.$ {\em Expert Annotation}, and $ii.$ {\em Annotation Aggregation}. The former one refers to the main annotation task, where a set of trajectories are assigned to human experts for annotation. The second component is to finalize the annotations for each trajectory by aggregating existing annotations provided by different human experts and land on a consensus.

\input{ExpertAnnotation}

\input{Aggregation}

%% file: ExpertAnnotation.tex
\subsection{Expert Annotation}
\label{sec:expAnnot}

This component consists of three principled sub-procedures: $i.$ Annotation Portal Preparation, $ii.$ Human Expert Preparation, and $iii.$ Annotation Process. In the following, we describe each sub-procedure in details. 

\begin{figure*}[ht]  
  \centering
  \includegraphics[scale=0.42]{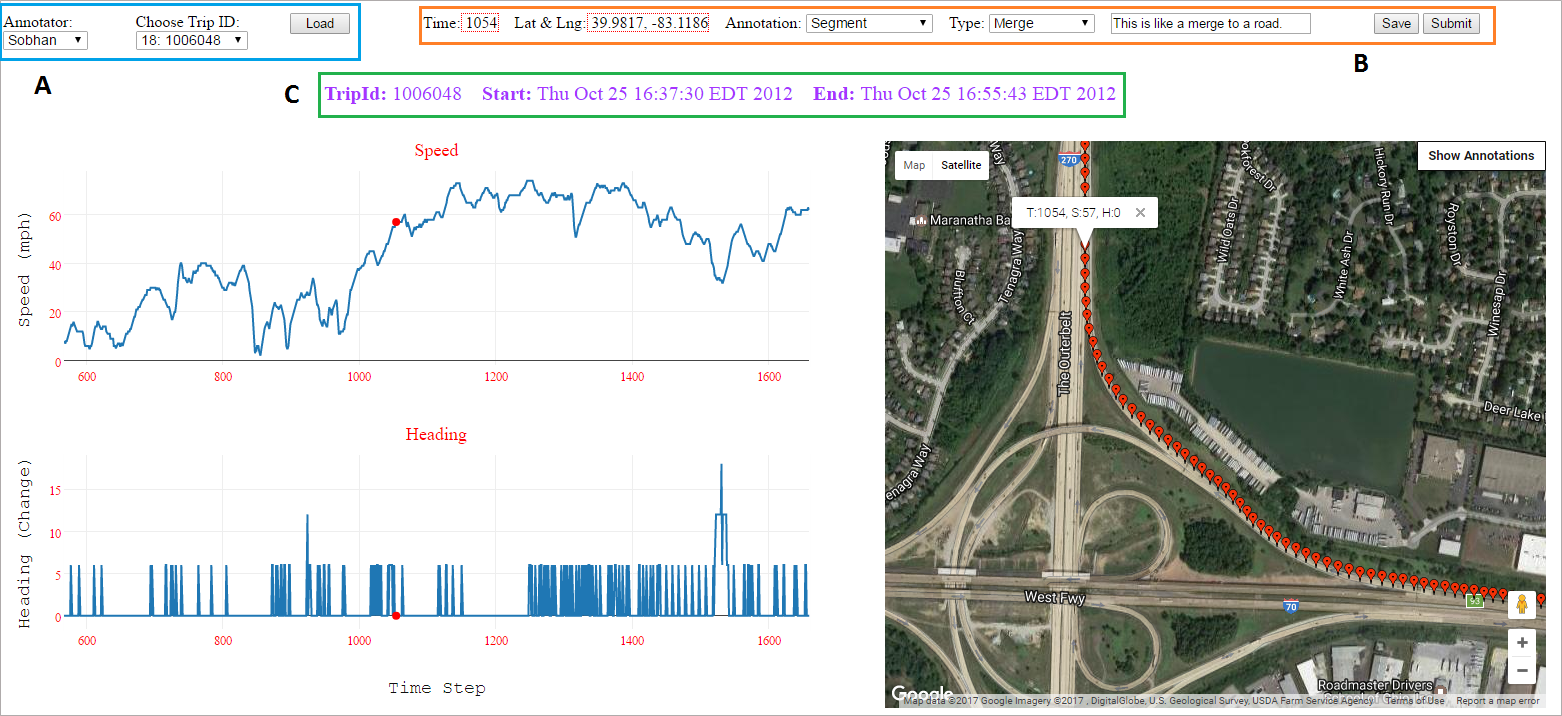}  
  \caption{\scriptsize {Trajectory Annotation Portal. The trajectory under investigation is represented on a map (right). The speed and heading change profiles are also reported (left). An annotator can select a point on any of these three diagrams and annotate it by using controls in Box $B$. The first step is to choose if the point represents the end of a segment or not (Segment, Maybe-Segment, or Non-Segment). The next step is to specify the type of the pattern (e.g. turn, speed-up, exit, etc.).}}
  \label{fig:portal}
\end{figure*}

\subsubsection{Preparing Annotation Portal}
The user interface for our annotation process is a web-based portal which represents a set of information for each trajectory to enable experts make decisions on annotations. Our interactive portal is implemented in PHP and JavasScript. A screenshot of the portal is illustrated in Figure \ref{fig:portal}. Different components of this portal are as follows.
\begin{itemize}[leftmargin=*]
    \item \textbf{Annotator Data Box} (Box $A$ in Figure \ref{fig:portal}) provides basic information such as annotator's name and also the list of trajectories (trips) assigned to the annotator. 
    \item \textbf{Annotation Box} (Box $B$ in Figure \ref{fig:portal}) provides functionalities to enter annotations. For this aim, the annotator should first opt the annotation type (Segment, Maybe-Segment, or Non-Segment) and then select a type (e.g., speed-up, turn, exit, merge, etc.).
    \item \textbf{Trajectory Time Box} (Box $C$ in Figure \ref{fig:portal}) provides information about the start and end time of the trajectory.
    \item \textbf{Speed Profile} is the time series of speed values throughout the trajectory measured in {\em miles per hour (mph)}. The X and Y axes provide time and speed data, respectively.  
    \item \textbf{Heading Change Profile} is the time series of direction change in direction of the moving vehicle. The change of heading $H_t$ for timestamp $t$ is calculated by $|Heading_t - Heading_{t-1}|$, where $Heading_t$ shows the heading of vehicle for timestamp $t$ and $|.|$ gives the absolute value. The X and Y axes represent the time and change in heading values, respectively. We employ {\em Plot.Ly} API\footnote{\scriptsize https://plot.ly/} to construct both speed and head change diagrams. 
    \item \textbf{Trajectory On Map}: We illustrate location-oriented information of the trajectory on a geographical map. We employ {\em Google Maps} API\footnote{\scriptsize https://developers.google.com/maps/} to visualize maps. 
\end{itemize}

The annotation process can be described as follows.
\begin{itemize}[leftmargin=*]
    \item [i. ] The annotator choose a trip ID and hit the load button. 
    \item [ii. ] Following the trajectory on any of three diagrams, the annotator may decide to choose a point as the end point of segment (pattern) by clicking on that. 
    \item [iii. ] To annotate the selected point (current segment), the annotator should choose the annotation type which is one of ``Segment'', ``Maybe-Segment'', or ``Non-Segment''. The ``Segment'' choice is picked when the expert is completely sure about a point to be the end point of segment. For example, in Figure \ref{fig:portal} we observe that the selected point on map is the end point of a segment which is a merge into a highway. However, in the case of uncertainty about the segment, an annotator can choose the ``Maybe-Segment'' option. For instance, there may be a pattern which is like a turn, but the annotator is not sure about it. 
    Another case which may lead to choose the ``Maybe-Segment'' is where we have a sub-pattern within a main pattern. For example, we may have a {\em slow-down} within a {\em turn} (see Figure \ref{fig:strict_aggregation}). In such a case, the annotator can annotate the slow-down as ``Maybe-Segment'' and the turn as ``Segment''. 
    Finally, if the annotator wants to undo her previously assigned annotations, she can choose the same point and assign the ``Non-Segment''.
    \item [iv. ] Once the annotation type is chosen, the next step is to decide about the type of segment (pattern). The list of our observed patterns are described in Table \ref{tab:pattern_types}. 
    \item [v. ] The annotator proceeds the annotation process by selecting another point of the trajectory and iterating back to step $ii$. Once the annotator reached the end of the trajectory, she should submit her annotations. 
\end{itemize}

We leverage some functionalities of {\sc Google Maps} API to synchronize all three diagrams and enable coordinated views. This makes the annotation task more straightforward for human annotators. When an annotator moves the mouse on the map, she notices a red marker on Speed and Heading diagrams designating the same point. Moreover, the annotator can click on {\em Show Annotation} button anytime during the annotation process to track her previously marked annotations. Annotators can also benefit from the functionalities of {\sc Plot.Ly} API to zoom-in into speed or heading diagrams in order to make wiser annotations. 

\begin{table*}
    \scriptsize
    \caption{\small Description of different types of segments (patterns), which can be used as labels for identified segments.}
    \centering
    \begin{tabular}{| c | c | c |}
        \hline 
        \textbf{Pattern} & \textbf{Description} & \textbf{Visual Demonstration}\\
        \hline
        \hline
        Exit & Car exists a highway/road & \raisebox{-.5\height}{\includegraphics[scale=0.14]{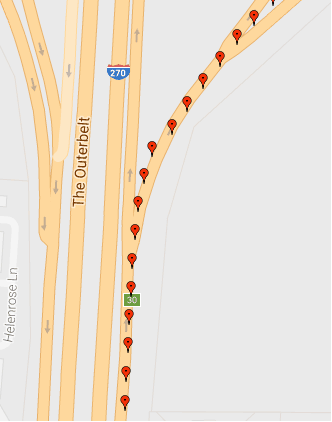}}\\
        \hline
        Merge & Car merges into a highway/road, in this case we have changes in speed value. & \raisebox{-.5\height}{\includegraphics[scale=0.14]{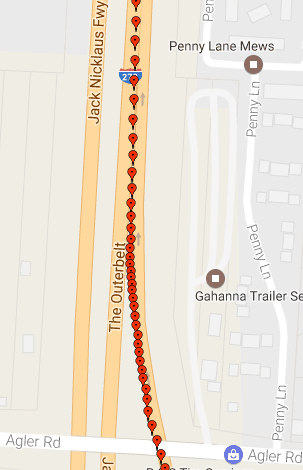}}\\
        \hline
        Exit-Merge & Car exists a highway/road and merges into another one. & \raisebox{-.5\height}{\includegraphics[scale=0.14]{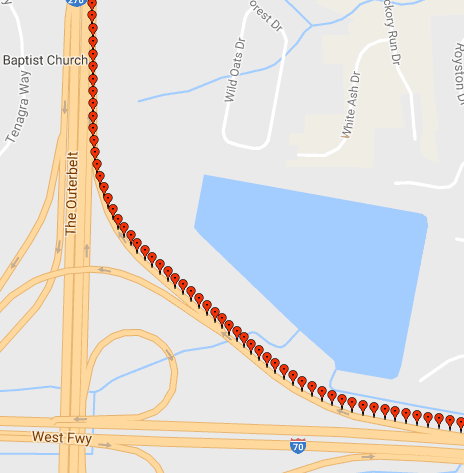}}\\
        \hline
        Loop & Car takes the ramp and makes a complete loop to merge into a highway/road. & \raisebox{-.5\height}{\includegraphics[scale=0.14]{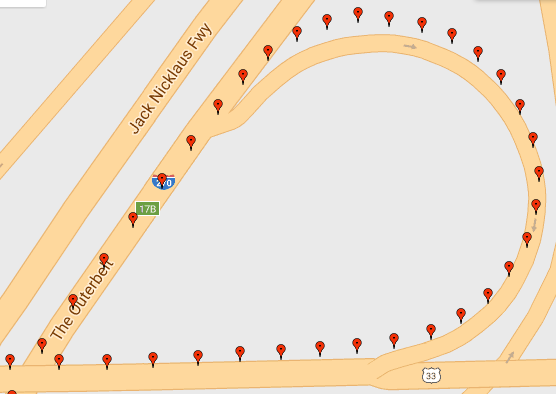}}\\   
        \hline
        Turn & \begin{tabular}{@{}c@{}} Car makes a turn which has significant effect \\  in change of direction of moving vehicle. \end{tabular} & \raisebox{-.5\height}{\includegraphics[scale=0.14]{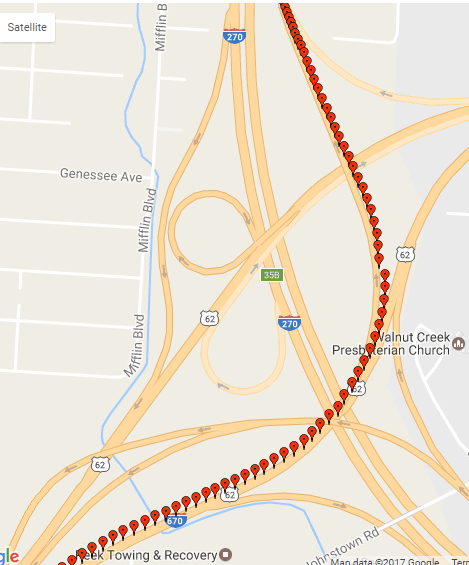}}\\
        \hline
        Smooth-Turn & \begin{tabular}{@{}c@{}} Car makes a smooth turn which has some effects \\ in change of direction of moving vehicle.\end{tabular} & \raisebox{-.5\height}{\includegraphics[scale=0.2]{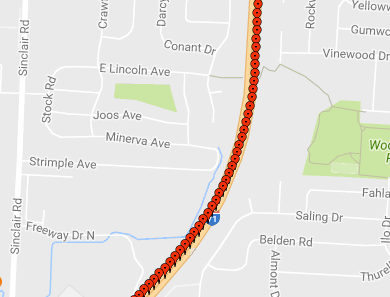}}\\
        \hline
        Left-Turn & Car makes a left turn in an intersection. & \raisebox{-.5\height}{\includegraphics[scale=0.18]{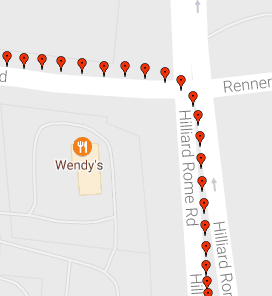}}\\
        \hline
        Right-Turn & Car makes a right turn in an intersection. & \raisebox{-.5\height}{\includegraphics[scale=0.23]{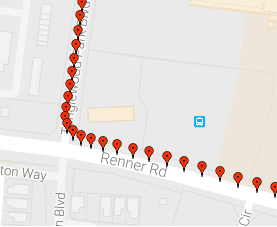}}\\        
        \hline 
        Jiggling & \begin{tabular}{@{}c@{}}Car keeps moving to left and right which has continues effect in change of \\ direction of vehicle, as it is observable from heading change values.\end{tabular} & \raisebox{-.5\height}{\includegraphics[scale=0.23]{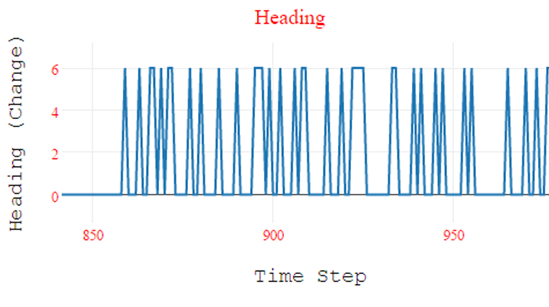}}\\            
        \hline
        Speed-Up & Car speeds up to merge a highway, after traffic light, after stop sign, etc. & \raisebox{-.5\height}{\includegraphics[scale=0.35]{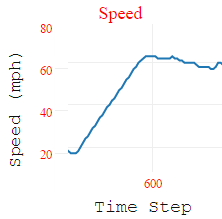}}\\
        \hline
        Slow-Down & \begin{tabular}{@{}c@{}} Car slows down because of traffic congestion, traffic light, \\  stop sign, intersection, etc. \end{tabular} & \raisebox{-.5\height}{\includegraphics[scale=0.32]{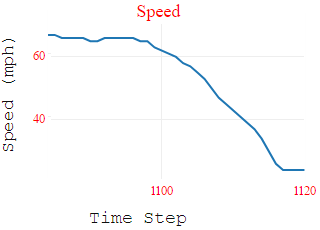}}\\
        \hline
        Traffic-Light & \begin{tabular}{@{}c@{}} Car stops behind a traffic light for a while, in such case, \\ we can see the speed is zero or has some minor changes. \end{tabular}&  \raisebox{-.5\height}{\includegraphics[scale=0.33]{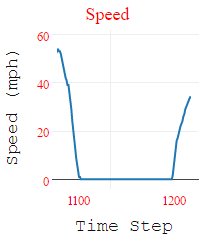}}\\        
        \hline      
    \end{tabular}
    \label{tab:pattern_types}
\end{table*}

\subsubsection{Preparing Human Experts}
In order to prepare human experts for annotation, we perform following steps:
\begin{itemize}
    \item [I. ] \textbf{Preparing a Tutorial}: As tutorial, we prepared two set of resources: $.i$ Instructive document which contains the description of the task and the annotation portal, and $ii.$ A short video clip to describe the task and annotate a sample trajectory by using the portal.    
    \item [II. ] \textbf{Running a Pilot Phase}: We ran a pilot phase of annotation where we assigned two trajectories to each annotator. The goal of running this phase was to give this opportunity to the annotators to work with the portal, comprehend the task, digest the process, and reduce the potential errors for the expert-annotation phase. 
\end{itemize}

\subsubsection{Main Annotation}
\label{subsec:main_annot}
In this step, we assign a set of trajectories to each annotator. We randomly distribute trajectories among annotators, such that each trajectory is assigned to exactly two different annotators.  
The number of assigned trajectories was not the same for all annotators, and it was determined based on a combination of few constraints such as {\em time}, {\em availability}, and {\em expertise}. 
Among annotators, we had four students, two researchers and a car insurance expert.

%% file: Aggregation.tex
\subsection{Annotation Aggregation}
\label{sec:aggr}
The annotation aggregation is to finalize subjective decisions and generate the final dataset of annotated car trajectories. The aggregation task can be regarded in terms of three following important decision items: to accept, refine or reject an annotation.


The most important challenge in aggregation is the {\em subjectivity} of annotation, where there may not be some strong agreements among annotators for some cases. In order to deal with such subjectivity, we perform the aggregation in two different phases: 1) Strict aggregation, and 2) Easy aggregation. We describe both later in this section.
Moreover, we leverage a set of heuristic-based generated annotations along with human expert annotations to be used as input for aggregation process, as it will be described later in this section.
Also, to successfully accomplish the aggregation task, we designed an {\em Aggregation Portal} which we describe next.

\subsubsection{Aggregation Portal}
The desiderata for the aggregation task are as follows.
\begin{itemize}[leftmargin=*]
    \item Trajectory data: We need the trajectory data in order to finalize the annotations. Such data includes speed profile, changes of heading, and also map representation. 
    \item Annotations by experts: This is the main input for aggregation. We need to have the annotations by different experts to be appropriately represented for analysis and decision making.
    \item Decision making features in order to land on an annotation aggregation.
\end{itemize}

\begin{figure*}[ht]  
  \centering
  \includegraphics[scale=0.42]{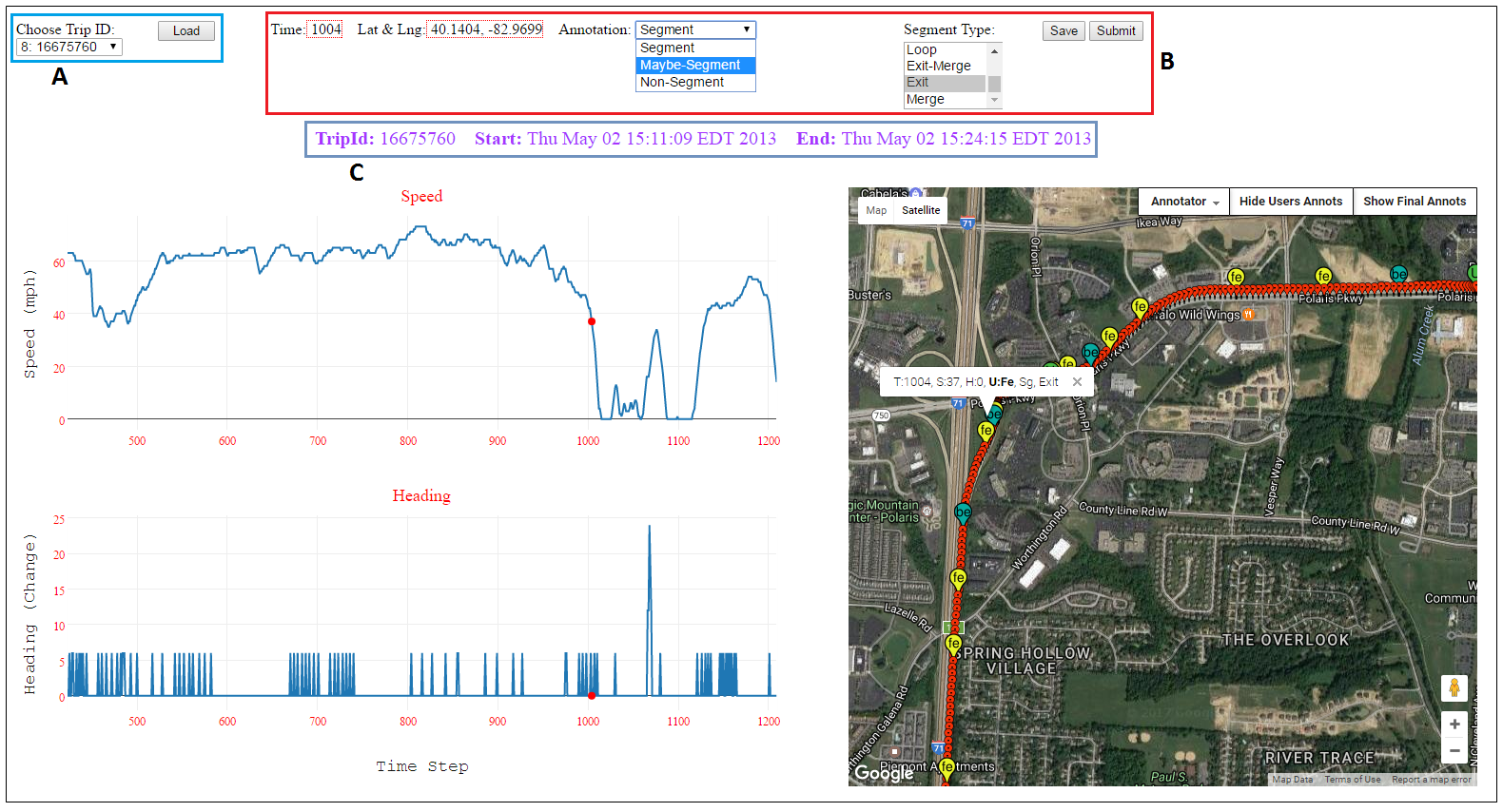}  
  \caption{\scriptsize {Annotation Aggregation Portal.. The trajectory is illustrated on map (right). The Speed and Change in Heading profiles are illustrated (left). The aggregation expert employs controls on the map to show or hide the annotations. To finalize annotations, the aggregator may decide to keep, refine, or ignore annotations. In order to make a decision about acceptance/refinement, the aggregator may choose a point on either of three diagrams, and then make a decision by using controls in Box $B$.}}
  \label{fig:aggregation}
\end{figure*}

To reach the above desiderata, we designed a new portal akin to the one for the expert annotation phase (Section \ref{sec:expAnnot}.1) with extra features. A snapshot of this portal is depicted in Figure \ref{fig:aggregation}. The aggregation portal is composed of following components:

\begin{itemize}[leftmargin=*]
    \item \textbf{Trajectory Selection Box} (Box $A$ in Figure \ref{fig:aggregation}) enables the aggregator expert to pick a trajectory for annotation.
    \item \textbf{Annotation Box} (Box $B$ in Figure \ref{fig:aggregation}) is employed to finalize the annotations for a trajectory. The controls in this box are almost identical to ones in the annotation portal (Figure \ref{fig:portal}). However, the aggregator expert is able to choose multiple types to specify the segment type. This feature is considered to deal with subjectivity of the task.
    \item \textbf{Trajectory Time Box} (Box $C$ in Figure \ref{fig:aggregation}) is used to show the real start and end time of the selected trajectory.
    \item \textbf{Speed Profile} represents the time series of speed values for the selected trajectory. 
    \item \textbf{Heading Change Profile} represents the profile of the changes in heading values for the selected trajectory.
    \item \textbf{Trajectory On Map}: The trajectory is represented on a geographical map and is enriched with some controls for the annotation task. Additional control for the aggregator is ``Show/Hide Users Annots'' and ``Annotator'' buttons. The first button is to show or hide the annotations which are made by experts during the expert-annotation phase (Section \ref{sec:expAnnot}.3), and the second button is to choose the annotators who want to show or hide their annotations, instead of showing or hiding all annotators data by using the first button. 
\end{itemize}

\subsubsection{Strict Aggregation}
\label{sec:strict_aggregation_phase}
\begin{figure*}[t]
    \centering
    \begin{subfigure}[b]{0.44\textwidth}
            \includegraphics[width=\linewidth]{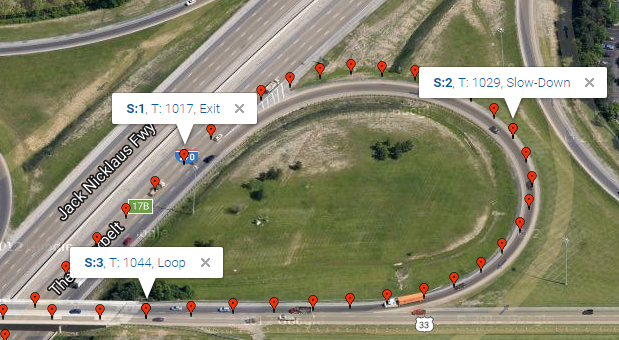}
            \caption{Strict Aggregation}
            \label{fig:strict_aggregation}
    \end{subfigure}\hspace{0.08\textwidth}
    \begin{subfigure}[b]{0.44\textwidth}
            \includegraphics[width=\linewidth]{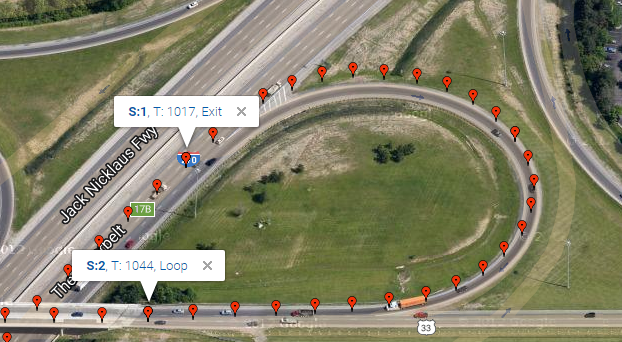}
            \caption{Easy Aggregation}
            \label{fig:easy_aggregation}
    \end{subfigure}%
    \caption{\small {Difference between \textbf{strict} (a) and \textbf{easy} (b) aggregation: In the first case, we consider both {\em intermediate} and {\em main} segments (patterns), while, for the second one, we just consider the main segments. Intermediate segments will be annotated as {\tt MAYBE-SEGMENT} and the main segments as {\tt SEGMENT}. Examples of {\em main} segments are S1 and S3 in (a) and an example of {\em intermediate} segment is S2 in (a).}}
    \label{fig:strict_easy_compare}
\end{figure*}

During the strict aggregation phase, we focus on the subjectivity of the aggregation task and opt to maximize the usage of users' input by considering following principles: 

\begin{itemize}[leftmargin=*]
    \item \textbf{Catch any Existing Segment}: Different annotators may identify different sets of segments in a trajectory. By strict aggregation, we try to catch any existing segment. In other words, we aim to annotate both {\em main} and {\em intermediate} segments (patterns). Main segments are those patterns which are more obvious, like an entire loop (Table \ref{tab:pattern_types}), and intermediate segments are those patterns which occur inside a main segment, like the slow-down within a loop. Figure \ref{fig:strict_aggregation} shows an example of strict aggregation where an intermediate segment (S2) is annotated along with two main segments (S1, and S3). Main segments will be annotated as {\tt SEGMENT} and intermediate segments as {\tt MAYBE-SEGMENT}.
    \item \textbf{Use Strict Thresholds}: When an annotator is looking for patterns like speed-up, slow-down, smooth-turn, or jiggling, the question is how to identify them? What are the requirements to catch a segment and label that with either of the mentioned types? The {\em answer} to such question is where we define the concept of strictness in contrast with easiness. During the strict aggregation, any change of the speed by at-least \underline{5 mph} is a significant change and the related segment will be annotated by speed-up or slow-down. Moreover, any continues change in heading values for \underline{five consecutive seconds}\footnote{\scriptsize As we may have GPS drifts in data leading to erroneous heading values, we catch significant changes in heading values based on a continuous pattern, instead of a single change.} is a significant change in heading and related pattern will be annotated as smooth-turn or jiggling. We found these numbers in trial-and-error experiments and based on the expertise of some of annotators. 
    \item \textbf{Assign all Relevant Segment Types}: Unlike the expert-annotation phase, the aggregator expert in aggregation phase can assign as many segment types as she thinks are relevant. As a common example, when a car exits a highway, a significant slow-down shall often happen. In such a case, the aggregator expert can assign both {\tt EXIT} and {\tt SLOW-DOWN} to the related segment.
\end{itemize}

\subsubsection{Easy Aggregation}
\label{sec:easy_aggregation_phase}
The easy aggregation is defined in contrast with strict aggregation, which is based on the following important principles.

\begin{itemize}[leftmargin=*]
    \item \textbf{Annotate Main Segments}: Unlike the strict aggregation, the aggregator expert will just annotate the main segments and will ignore the intermediate ones during the easy-aggregation. An example of easy aggregation is shown in Figure \ref{fig:easy_aggregation}, where both S1 and S2 are main segments (annotated as {\tt SEGMENT}) and the slow down pattern (S2 in Figure \ref{fig:strict_aggregation}) is not selected by the aggregator. 
    \item \textbf{Use Relaxed Thresholds}: The second important difference between strict and easy aggregation is the selection of thresholds which will be used to catch patterns like speed-up, slow-down, smooth-turn, and jiggling. In case of easy aggregation, we use some relaxed version of the previously defined thresholds which are as follows: for speed, a change by at-least \underline{10 mph} is a significant change. Moreover, for heading, we look for a consecutive series of changes which last for at-least \underline{10 seconds}. By using these relaxed thresholds, we expect to see less patterns of above mentioned types. 
    \item \textbf{Assign all Relevant Segment Types}: Similar to strict aggregation, the aggregator expert will assign all relevant segment types to an identified segment.
\end{itemize}

\subsubsection{Heuristic-Based Annotation}
\label{sec:heuristic_annot}
In order to deal with subjectivity and also to make sure to have a comprehensive set of potential annotations prior to aggregation, we leverage a set of heuristically generated annotation, in addition to human expert annotations (Section \ref{sec:expAnnot}), to be used as main input for aggregation process. 
Being provided alongside human annotations, it enables experts to get to a consensus among annotators.
The heuristic annotations are generated by an algorithm called {\sc AutoAnn}. 
The results of the algorithm is a second-order guide to the aggregator to decide on the best segments.
{\sc AutoAnn} heuristics are based on obvious patterns we observed during initial investigations. {\sc AutoAnn}'s logic is based on speed-based, position-based and heading-based heuristics. We stress that employed heuristics are very simplistic and cannot replace the human power of decision making. Nonetheless, the benefit of an automated annotator is two-folded.
 
{\sc AutoAnn} scans each single point in a trajectory (in the ascending order of timestamps) and makes comparisons with $k$ neighbor points. The $k$ parameter depends on the temporal precision of records in the dataset. Based on our data observations, we experimentally set $k = 5$. Given the temporal precision of our dataset, i.e., seconds, $k=5$ implies that we analyze behaviors in a $5$-second time window. For a given point $p$ and its $k$ neighbor points, {\sc AutoAnn} preforms following heuristics.

\begin{itemize}[leftmargin=*]
    \item Speed-wise, if $p$ falls into a local maxima among it's $k$ neighbors (i.e., $k$ previous and $k$ subsequent points), then $p$ marks the end of a ``speed-up'' segment.
    \item Similarly, if $p$ falls into a local minima, $p$ marks the end of a ``slow-down'' segment.
    \item If $p$ holds as the end of a ``slowdown'' and the speed at $p$ is lower than a threshold, then $p$ marks the end of a ``traffic-jam'' segment. Based on our data observation, we experimentally set the low speed threshed to $9$ miles per hour. The intuition is that a {\em serious} slow-down should be due to a traffic jam. Normal slowdowns may occur due a slight turn or sun rays where the speed does not get lower than $9$ miles per hour.
    \item If the speed of all $k$ previous points of $p$ are lower than the low speed threshold and the speed of the immediate point after $p$ is larger then the low speed threshed, the $p$ marks the end of a ``traffic-light'' segment.
    \item We define {\em heading-change} as the first-order derivative of $p$'s and $p'$'s heading values where $p'$ is the immediate point after $p$. To simplify computation, we replace {\em derivatives} with the mathematical subtraction. The point $p$ marks the end of a ``turn'' segment iff it satisfies following conditions: $(i)$ The heading-change of $k$ neighbors of $p$ are zero and $(ii)$ The heading-change at $p$ is larger than a threshold. We experimentally set the high heading change threshold to $15$. 
    \item If the position of the point $p$ is merely identical with the point $p'$ which happened before $p$, then $p'$ marks the beginning and $p$ marks the end of a ``loop'' segment. The intuition is that when a driver performs a loop (for instance through a highway exit), the first and last locations of the loop are identical with a tiny error. However, this position-based heuristic cannot be applied on our data because the same situation is reported when the car is still. A part of our future work is to make this heuristic more sophisticated. One potential extension is to consider speed, the heuristic which seems to be valid is $p$'s speed is larger than a minimum threshold.
\end{itemize}

On our data, {\sc AutoAnn} generates 2418 segments among which 10\% are speed-ups, 59\% are slow-downs, 6\% are traffic-jams, 20\% are traffic-lights and 5\% are turns. Of course {\sc AutoAnn} introduces many false positives and false negatives. 

%% file: Dataset.tex
\section{Trajectory Dataset}
\label{sec:data}
We employ a subset of a large-scale, real-world dataset of personal car trajectories, collected using highly accurate devices connected to On Board Diagnostic (OBD-II) port of vehicles. The data is collected with a consistent sampling rate of one second, whereas most of the existing public datasets provide inconsistent or lower rates for data collection process \cite{nycTaxi, geolife-gps-trajectory-dataset-user-guide, moreira2013predicting}. For instance, driving data is collected every other 15 seconds in \cite{moreira2013predicting} and data collection rate is varying between 1 and 5 seconds in \cite{geolife-gps-trajectory-dataset-user-guide}.

The set of data items which were collected every second includes {\em speed}, {\em acceleration}, {\em heading}, {\em GPS coordinates}, and {\em time stamp}. We consider following criteria to obtain a subset of the dataset for annotation.
\begin{itemize}[leftmargin=*]
    \item [--] Remove trajectories with missing point(s).
    \item [--] Keep trajectories with diverse set of drivers.
    \item [--] Keep trajectories with a fair diversity of traffic condition.
    \item [--] Keep trajectories with a fair diversity of highway and metropolitan coverage.
\end{itemize}

Based on the above criteria, we obtained a sample of the trajectory dataset which is summarized in Table \ref{tab:dataset_fact}. The sampled dataset covers a fair amount of driving data covering about 13 hours of drive. Moreover, the diversity of data is another strength factor of this sampled dataset which is measured as the number of drivers who are involved in generation of the driving data.
It is also worth mentioning that most of the trajectories in the sampled dataset occurred between 3pm and 7pm, thus they cover a fair amount of both {\em rush hour} and {\em normal traffic condition}. 

\begin{table*}[t]
    \centering
    \small
    \caption{ Some statistical facts about the sampled {\em Trajectory Dataset} for annotation}
    \begin{tabular}{| c | c | c | c | c | c |}
        \hline
        \begin{tabular}{@{}c@{}} \textbf{Number of} \\ \textbf{Trajectories} \end{tabular} & \begin{tabular}{@{}c@{}} \textbf{Number of} \\ \textbf{Drivers} \end{tabular} &    \begin{tabular}{@{}c@{}} \textbf{Total} \\ \textbf{Driving Time} \end{tabular} & \begin{tabular}{@{}c@{}} \textbf{Total Driving} \\ \textbf{in Highway} \end{tabular} & \begin{tabular}{@{}c@{}} \textbf{Total Driving} \\ \textbf{in City} \end{tabular} & \begin{tabular}{@{}c@{}} \textbf{Average} \\ \textbf{Trajectory Length} \end{tabular}\\ [0.6ex]
        \hline
        50 & 19 & 13.3 Hours & 12.3 Hours & 1 Hours & 16 Minutes \\
        \hline
    \end{tabular}
    \label{tab:dataset_fact}
\end{table*}

%

%% file: Results.tex
\section{Annotation Results}
\label{sec:res}
In this section we first provide some detailed description of {\em inter-annotator agreement} analysis. Then, we present the results for different annotation phases in terms of {\em Statistical Facts}, {\em Distribution of Annotation Types}, {\em Distribution of Segment Types}, and {\em Agreement Analysis}.

\subsection{Inter-Annotator Agreement Analysis}
\label{sec:inter_annotator_agreement}
One of the crucial experiments based on the results of annotation is to see how much agreement exist between the annotators. For this aim, we conduct an {\em inter-annotator agreement} analysis. We formulate this task as follows: 
suppose for a trajectory $T$, we have two annotators A and B who provided two different sets of annotations $Annot_A = \{a_{A1}, a_{A2}, \dots,$ $ a_{An}\}$ and $Annot_B = \{a_{B1},a_{B2}, \dots, a_{Bm}\}$, where each $a_{Xi}$ is an annotation tuple specifying the exact point of annotation in $T$ as well as types of annotation and segment. Lets assume each annotation tuple $a_{Xi}$ is like a label {\em Yes} for corresponding data point of $T$. Similarly, we can assume each data point of $T$ which we don't have any annotation for that is labeled as {\em No}. In this way, Table \ref{tab:annotator_judgments} simplifies the agreement between A and B. In this table, the non-negative integer $a$ shows the intersection between $Annot_A$ and $Annot_B$. 
Moreover, the non-negative integer $d$ shows for how many data points of trajectory $T$, both annotators A and B decided to not assign any annotation. Similarly, $b$ ($c$) shows the number of data points of $T$ which for them we have annotations in $Annot_A$ ($Annot_B$), but have no annotation in $Annot_B$ ($Annot_A$). 

We use two different measures to obtain correlation between $Annot_A$ and $Annot_B$, {\em Cohen's Kappa} and {\em Overlap}.
Cohen's Kappa is formulated by Equation \ref{eq:cohens_kappa}, regarding its original formulation as proposed by \cite{cohen1960coefficient} and information in Table \ref{tab:annotator_judgments}:

\begin{equation}\label{eq:cohens_kappa}
    \kappa = \frac{p_o - p_e}{1 - p_e}
\end{equation}
        
 where $\pmb{p_o} = \frac{a+d}{a+b+c+d}$, $\pmb{p_e} = p_{Yes} + p_{No}$, $\pmb{p_{Yes}} = \frac{(a+b)(a+c)}{(a+b+c+d)^2}$, and $\pmb{p_{No}} = \frac{(c+d)(b+d)}{(a+b+c+d)^2}$. Note that by using this measure, we consider the possibility of the agreement occurring by chance. 
 Also, based on Table \ref{tab:annotator_judgments}, we define the Overlap function by Equation \ref{eq:overlap}:

\begin{equation}\label{eq:overlap}
    Overlap = \frac{a}{a + b + c}
\end{equation}

As one can see, the Overlap is, in some sense, the fraction of intersection between $Annot_A$ and $Annot_B$.
Now, we describe how to obtain numbers $a$, $b$, $c$, and $d$ in Table \ref{tab:annotator_judgments}. Given annotation tuples $a_{Ai} \in Annot_A$ and $a_{Bj} \in Annot_B$, we calculate their Haversine distance\footnote{\scriptsize See https://en.wikipedia.org/wiki/Haversine\_formula.} using the latitude and longitude of each annotation tuple. Then, if their distance is less than a pre-defined threshold $\tau$, we say they are correlated and will increase $a$, otherwise, they are uncorrelated and will increase either $b$ (if failed to find a match for $a_{Ai}$) or $c$ (if failed to find a match for $a_{Bj}$). Finally, after finalizing the values of $a$, $b$, and $c$, we set $d = |T| - (a+b+c)$, where $|T|$ is the number of data points in trajectory $T$. 

\begin{table}
	\centering
	\caption{\small An example of annotation by two annotators. The annotation task is about labeling of the inputs as {\em Yes} or {\em No}. Numbers $a$, $b$, $c$, and $d$ show the agreements/disagreements between annotators A and B.}
	\begin{tabular}{| c | c | c | c |}
		\hline
		\multicolumn{2}{|c|}{\cellcolor{lightgray} { }}  & \multicolumn{2}{c|}{\textbf{B}} \\
		\cline{3-4}
		\multicolumn{2}{|c|}{\cellcolor{lightgray} { }}  & {\tt Yes} & {\tt No}\\		
		\hline
		\multirow{ 2}{*}{\textbf{A}} & {\tt Yes} & a & b \\ \cline{2-4} & {\tt No} & c & d \\
		\hline
	\end{tabular}
	\label{tab:annotator_judgments}
\end{table}

\vspace{0.1in}
\subsection{Results and Comparison}
\smallskip
\subsubsection{Statistical Facts}
In this section, we discuss some statistical facts about the annotation phases. Table \ref{tab:statistical_facts} provides some statistics about all three phases of annotation. We employed seven annotators for expert annotation phase and one aggregator annotator for each of the the strict and easy aggregation phases. 

We observe in Table \ref{tab:statistical_facts} that the most time-consuming phase was the strict aggregation and the number of extracted segments during this phase was also the largest among all phases. Another observation is on the average number of extracted segments for each trajectory, where this number is similar for expert-annotation and easy-aggregation phases, but almost half of the extracted segments for strict-aggregation. 
Moreover, the annotators and the aggregator of strict-aggregation spent the same amount of time to annotate a new segment (31 seconds). However, the aggregator provided more segments for each trajectory on average. This justifies the reason of spending about 25 minutes to segment a single trajectory by the aggregator, where annotator just spent about 10 minutes on average. 

\begin{table*}[t]
    \centering
    \small
    \caption{Some statistical facts about different phases of annotation}
    \begin{tabular}{c | c | c | c | c | c }
        \textbf{Annotation Phase} &
        \begin{tabular}{@{}c@{}} \textbf{Num of Segments} \\ \textbf{Specified By Expert(s)} \end{tabular} & \begin{tabular}{@{}c@{}} \textbf{Avg Num of Segments} \\ \textbf{for each Trajectory} \end{tabular} &    \begin{tabular}{@{}c@{}} \textbf{Total Active} \\ \textbf{Annotation Time} \end{tabular} & \begin{tabular}{@{}c@{}} \textbf{Avg Annotation Time} \\ \textbf{for a Single Trajectory} \end{tabular} & \begin{tabular}{@{}c@{}} \textbf{Avg Time to Specify} \\ \textbf{a Single Segment} \end{tabular}\\ [0.6ex]
        \hline
        \textbf{Expert Annotation} & 1,997 & 20 & 17 Hours & 10.3 Minutes & 31 Seconds \\
        \hline
        \textbf{Strict Aggregation} & 2,465 & 49 & 21 Hours & 25.2 Minutes & 31 Seconds \\
        \hline
        \textbf{Easy Aggregation} & 1,372 & 27 & 6.3 Hours & 7.5 Minutes & 15 Seconds \\
    \end{tabular}
    \label{tab:statistical_facts}
\end{table*}

\subsubsection{Distribution of Annotation Types}
\begin{figure}[t]  
  \centering
  \includegraphics[scale=0.4]{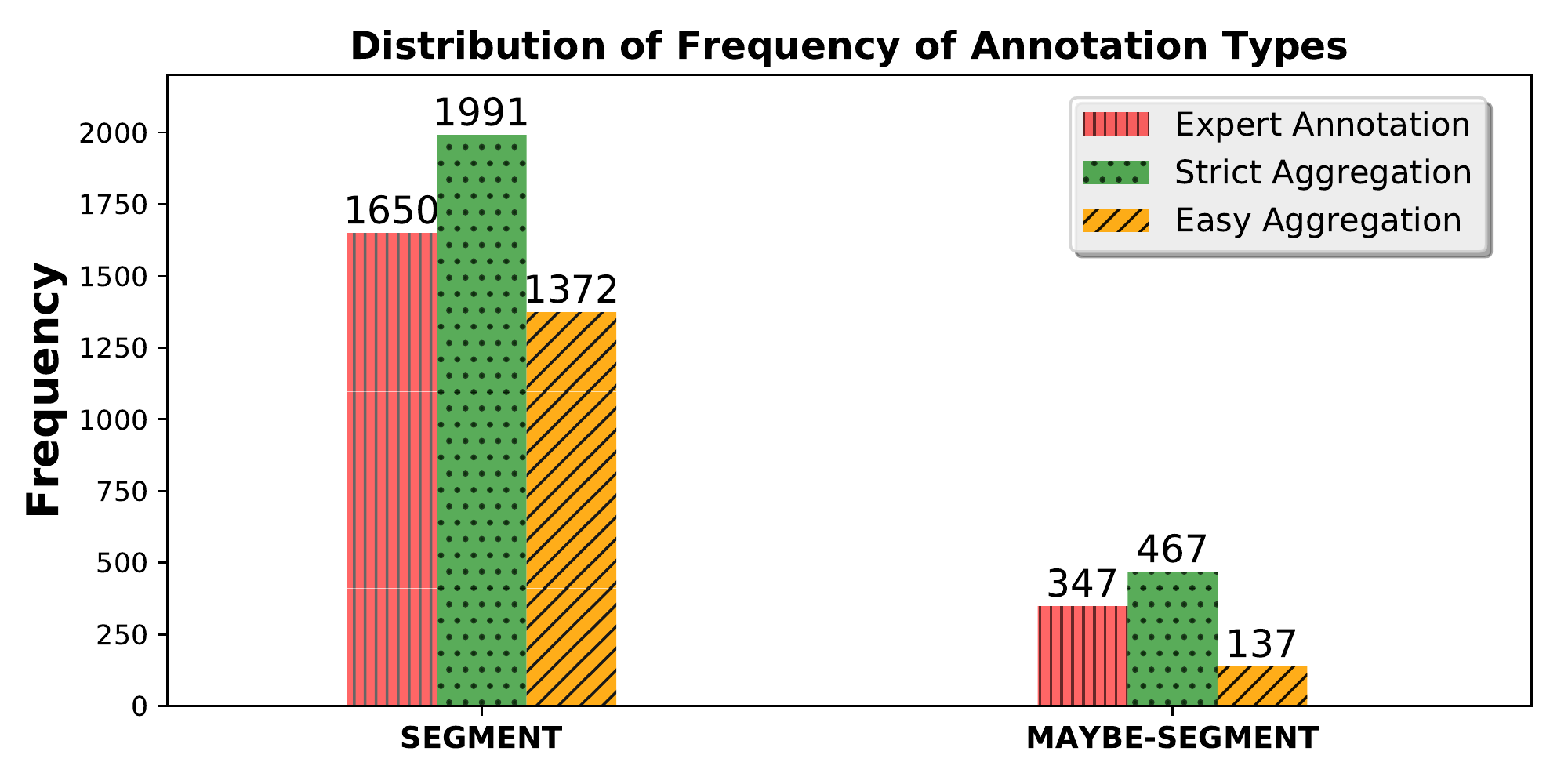}  
  \caption{Frequency distribution of Annotation Types ({\tt SEGMENT} and {\tt MAYBE-SEGMENT}). }
  \label{fig:annot_type_compare}
\end{figure}

Figure \ref{fig:annot_type_compare} shows the frequency distribution of annotation types for three phases of annotation. Recall that the possible annotation types are  {\tt SEGMENT} and {\tt MAYBE-SEGMENT}, where we mostly refer to former one as {\em main} and the later one as {\em intermediate} segment ( Section \ref{sec:strict_aggregation_phase}). 

In Figure \ref{fig:annot_type_compare}, we observe the same pattern of (relative) frequency for annotation types, where the majority of annotations for each phase is related to type {\tt SEGMENT}. However, an interesting observation is that for expert-annotation and strict-aggregation phases, we have about 80\% of annotations as {\tt SEGMENT}, while this number is about 90\% for easy-aggregation. This is interpreted as an outcome of using different principles for easy aggregation (as discussed in Section \ref{sec:easy_aggregation_phase}).   

\subsubsection{Distribution of Segment Types}
\begin{figure*}[t]  
  \centering
  \includegraphics[scale=0.5]{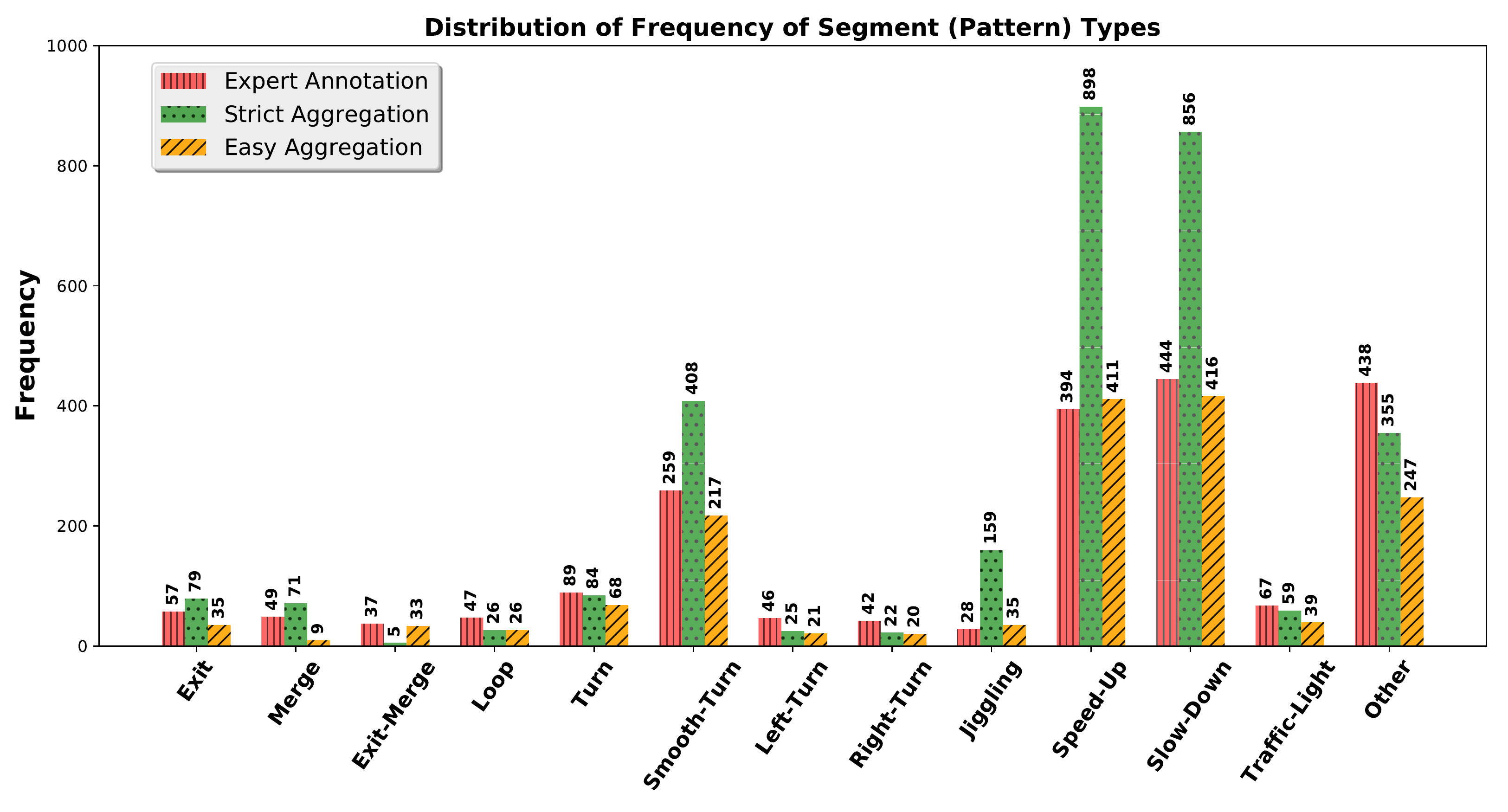}  
  \caption{Frequency distribution of Segment Types, as are listed in Table \ref{tab:pattern_types}, for different annotation phases. }
  \label{fig:segment_type_compare}
\end{figure*}

In this section, we discuss frequency distribution of different segment types (Figure \ref{fig:segment_type_compare}). We observe that the most common segment types are {\em speed-up}, {\em slow-down}, and {\em smooth-turn} which cover about 55\%, 70\%, and 66\% of annotations for expert-annotation, strict-aggregation, and easy-aggregation, respectively. 
Moreover, there is a tendency for the aggregator to assign less segments of the generic type {\em Other}, in comparison to annotators.
We also observe a balance between the frequency of correlated segment types. For example, there is some frequency-balance between {\em Exit} and {\em Merge}, {\em Left-Turn} and {\em Right-Turn}, or {\em Speed-Up} and {\em Slow-Down}, based on all three phases of annotation. 

\subsubsection{Agreement Analysis}

We report the agreement for each phase of the annotation as follows: 
\begin{itemize}[leftmargin=*]
    \item Expert-Annotation phase: given a distance threshold $\tau$, we first calculate the agreement for each pair of annotators, who annotated the same trajectory in the dataset (recall that each trajectory was assigned to two experts during the expert-annotation phase). Then, we obtain the average agreement between all pairs of annotators and report it as the agreement based on the distance threshold $\tau$.  
    \item Strict-Aggregation phase: given a distance threshold $\tau$ and a trajectory $T$, we first calculate the agreement between the aggregator of strict-aggregation phase and all annotators of $T$. Then, we obtain the average agreements for all pairs of $\langle super\-annotator,$ $expert \rangle$, and report it as the agreement based on distance threshold $\tau$. 
    \item Easy-Aggregation phase: the process for easy-aggregation is the same as for strict-aggregation phase, unless we calculate the agreement between annotations made by the aggregator of easy-aggregation phase and all other annotators of the same trajectory.
\end{itemize}

\begin{figure*}[ht]
    \centering
    \begin{subfigure}[b]{0.43\textwidth}
            \includegraphics[width=\linewidth]{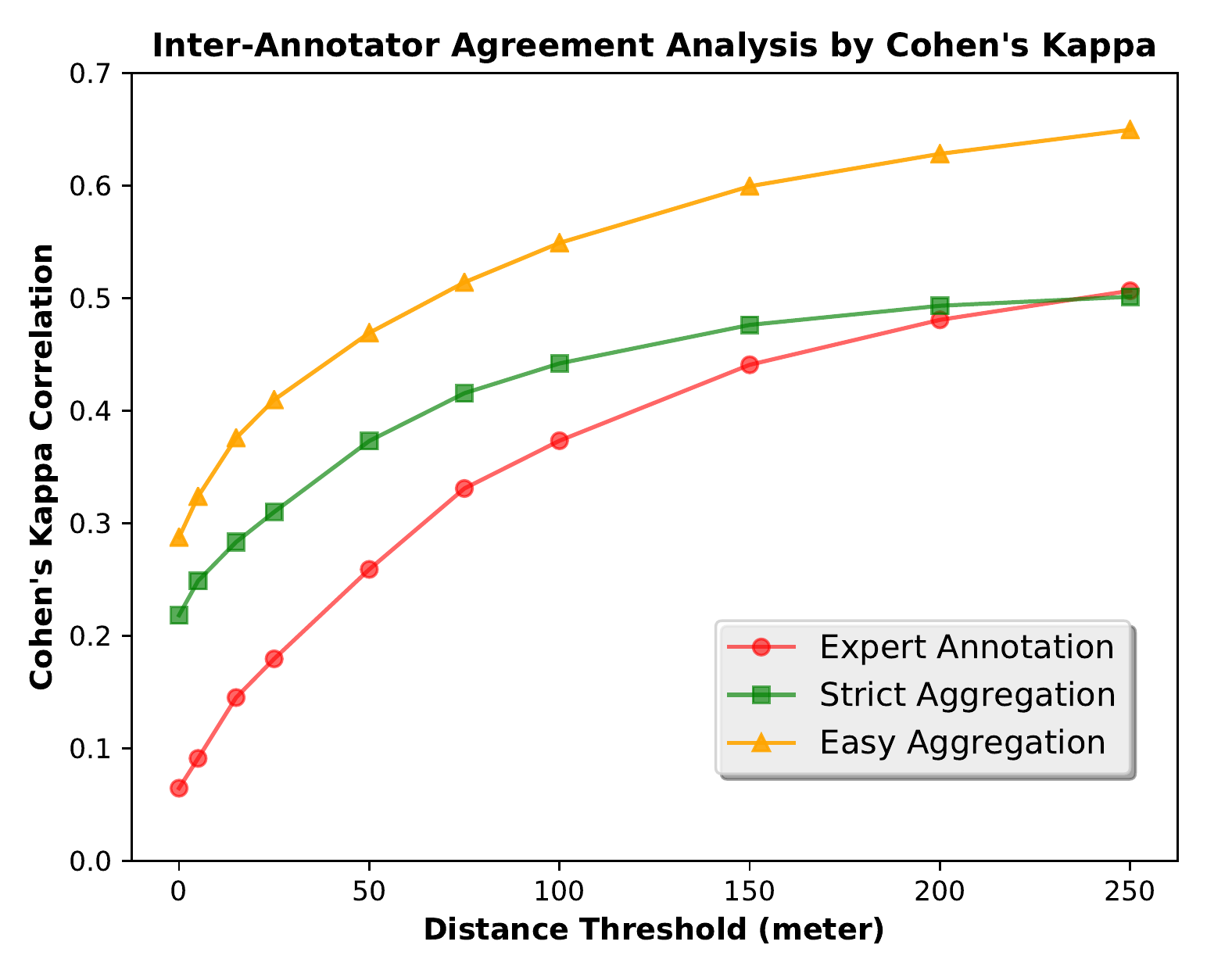}
            \caption{Cohen's Kappa}
            \label{fig:agreement_cohens}
    \end{subfigure}\hspace{0.08\textwidth}
    \begin{subfigure}[b]{0.43\textwidth}
            \includegraphics[width=\linewidth]{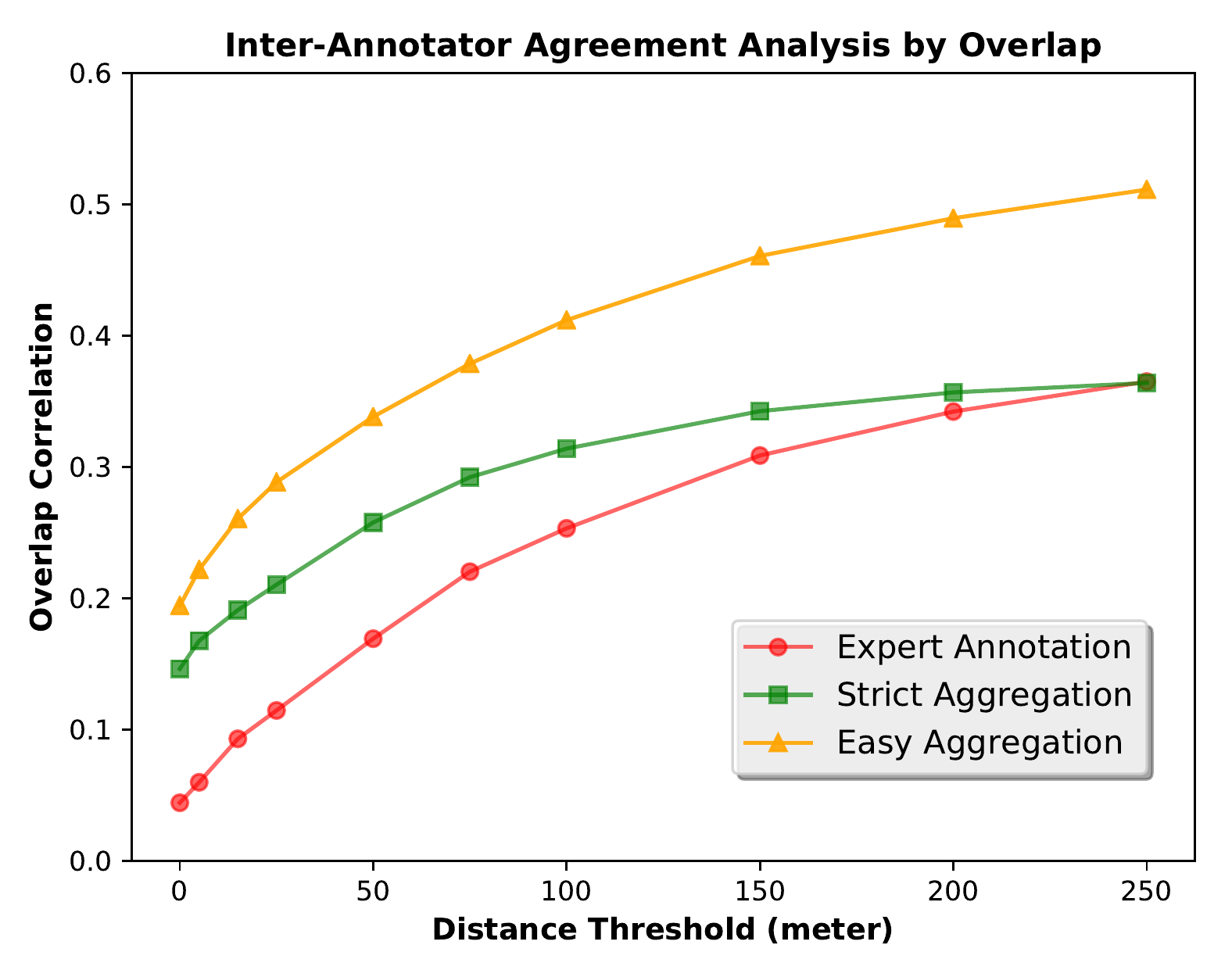}
            \caption{Overlap}
            \label{fig:agreement_overlap}
    \end{subfigure}%
    \caption{The result of intern-annotator agreement analysis based on (a) Cohen's Kappa and (b) Overlap metrics.}
    \label{fig:agreement_analysis}
\end{figure*}

Figure \ref{fig:agreement_cohens} shows the inter-annotator agreement analysis based on Cohen's Kappa, and Figure \ref{fig:agreement_overlap} shows the same analysis based on Overlap, for different distance thresholds. We observe that the largest agreement values based on both metrics are obtained for easy-aggregation phase. Also, the agreement for strict-aggregation is still larger than the agreement for expert-annotation. Based on the agreement analysis results, we can conclude following points.

\begin{itemize}[leftmargin=*]
    \renewcommand{\labelitemi}{\tiny$\blacksquare$}
    \item {\bf Subjectivity}. We observe that the disagreement between different annotators is large, even by employing a relaxed distance threshold like 200 meters or more. This justifies our claim on the subjectivity of trajectory annotation.
    \item {\bf Aggregation to Tackle Subjectivity}. We observe in Figures \ref{fig:agreement_cohens} and \ref{fig:agreement_overlap} that the aggregator utilizes the annotations of annotation phase, during both strict and easy aggregation phases. This enables a larger agreement for this two phases. 
    \item {\bf The closeness of Easy-aggregation to Expert-annotation}. Based upon Figures \ref{fig:agreement_cohens} and \ref{fig:agreement_overlap}, we conclude that there is larger agreement between results of easy-aggregation and expert-annotation phases, and this is due to employing some relaxed version of constraints during the easy aggregation, in comparison to strict aggregation phase. 
    \item {\bf Larger Agreement Values by Cohen's Kappa}. The other observation is that we obtain larger agreement values by using Cohen's Kappa. This is potentially related to using both labels {\em Yes} and {\em No} to obtain the correlation (agreement), while the Overlap exploits only the label {\em Yes} (Section \ref{sec:inter_annotator_agreement}). 
\end{itemize}

%% file: Output.tex
\vspace{0.2in}
\section{Dataset of Annotated Car Trajectories (DACT)}
\label{sec:output}
We describe the main outcome of this study which is a {\em Dataset of Annotated Car Trajectories}, abbreviated as {\sc DACT}. In this dataset, we have a collection of trajectories, where each trajectory is a time-ordered sequence of tuples. Each tuple consists of the following attributes: 
\begin{itemize}[leftmargin=*]
    \item [--] Trip ID: This is a unique identifier which is assigned to each trajectory. 
    \item [--] TimeStep: This is a positive integer which shows the index of current tuple in a sequence of tuples which constitute a trajectory.
    \item [--] TimeStamp: This is the sampling time of the current tuple which is reported in EDT (i.e. Eastern Daylight Time). 
    \item [--] Speed: This is the speed of vehicle in current moment, which is reported in {\em mph} (miles per hour). 
    \item [--] Acceleration: This is the acceleration of vehicle in current moment, which is reported in $m/s^2$, that is, meter per second squared. 
    \item [--] Heading: This shows the direction of moving vehicle at current moment which is a number between 0 and 359, where 0 means north and 180 means south. 
    \item [--] HeadingChange: This shows the change of heading of vehicle in comparison with the observed heading in the last timestep. By definition, the HeadingChange of the first timestep of a trajectory is set to $0$. 
    \item [--] Latitude: This shows the (GPS) latitude coordinate of the location of the vehicle in current moment.
    \item [--] Longitude: This shows the (GPS) longitude coordinate of the location of the vehicle in current moment.
    \item [--] Annotation: If there is any annotation for a tuple, it will be listed here. The possible values for this attribute are Segment, Maybe-Segment, and NULL (i.e., no annotation for a tuple). 
    \item [--] SegmentType: If there is any annotation for a tuple, the type(s) of segment will be listed here. The possible values for this attribute are NULL, Other, and any combination of segment types as are listed in Table \ref{tab:pattern_types}. 
\end{itemize}

As we had two phases of aggregation, i.e. strict and easy, the final dataset is provided in terms of two CSV files, one for each type of the aggregation. The columns of the CSV files are the same as attributes which we described above. The DACT dataset is available online at \url{ https://figshare.com/articles/dact\_dataset\_of\_annotated\_car\_trajectories/5005289}.

%% file: Conclusion.tex
\vspace{0.2in}
\section{Conclusion}
\label{sec:conc}
In this paper, we introduced a new trajectory annotation framework in order to annotate a dataset of personal car trajectories. Among annotation tasks, the trajectory annotation is challenging and needs more expertise and time to be accomplished appropriately. Consequently, the design of a framework for such task is complicated as well. 
Besides, one of the important challenges which we addressed during this study is the subjectivity of trajectory annotation. In order to deal with this challenge, we applied two separate phases of aggregation by using strict and relaxed (easy) constrains to finalize the annotations, and also leveraged the {\sc AutoAnn}, an annotator robot, to diminish the subjectivity.
Our inter-annotator agreement analysis shows that we increased the agreement from 44\% to about 60\% by aggregation based on Cohen's Kappa coefficient. 
Finally, the main output of this study which is a dataset of annotated car trajectories is made available and is publicly available for research projects. 

%% file: Acknowledgement.tex
\newpage
\section{Acknowledgement}
\label{sec:ack}
The authors would like to thank Dr. Placido A. Souza Neto and his research team at the Federal Institute of Rio Grande do Norte, and also the students of the Ohio State University for their priceless help and participation in the annotation project. We also want to thank the Nationwide Mutual Insurance Company for providing the valuable sources of data, which made this project possible.